\documentclass[preprint,5p,twocolumn]{elsarticle} 

\usepackage{amssymb}
\usepackage{dcolumn}
\usepackage{bm}
\usepackage{lineno}
\usepackage{upgreek}
\usepackage{multirow}
\usepackage{graphicx}
\usepackage{makecell}
\usepackage{amsmath}
\usepackage{array}

\setcitestyle{square,numbers,sort&compress}
\biboptions{numbers,sort&compress}

\journal{Infrared Physics and Technology}

\begin{document}

\begin{frontmatter}



\title{Polarization-insensitive dual-wavelength dispersion tunable metalens achieved by global modulation method}

\author [1]{Haohan Chen}
\author [1]{Qianbin Feng}
\author [1]{Jiepeng Wu}
\author [1]{Yanlin Zhu}
\author [1]{Hao Wang}
\author [1]{Qiang Li}
\author [1]{Lijun Wu \corref{cor1}}
\ead{ljwu@scnu.edu.cn}

\cortext[cor1]{Corresponding author}

\affiliation [1]{organization={Guangdong Provincial Key Laboratory of Nanophotonic Functional Materials and Devices, School of Information and Optoelectronic Science and Engineering, South China Normal University},
            city={Guangzhou 510006},
            country={China}}

\begin{abstract}
SF$_6$ is widely used as a gas-insulator in high-voltage power electrical system. Detecting SF$_6$ leaks using unmanned aerial vehicle (UAV)-based thermal cameras allows efficient large-scale inspections during routine maintenance. The emergence of lightweight metalenses can increase the endurance of UAVs. Simultaneously controlling dispersion and polarization properties in metalens is significant for thermal camera applications. However, via a propagation phase modulation method in which the phase is tuned locally, it is difficult and time-consuming to obtain enough different nanostructures to control multiwavelength independently while maintain the polarization-insensitive property. To this end, by using a global modulation method, a polarization-insensitive dual-wavelength achromatic and super-chromatic metalens are designed respectively. The working wavelength is set at 10.6 and 12 $\upmu \rm{m}$ to match the absorption peaks of SF$_6$ and one of its decompositions (SO$_2$F$_2$), respectively. According to the operating wavelengths, only the geometric parameters of two nanofins are required to be optimized (through genetic algorithm). Then they are superimposed on each other to form cross-shaped meta-atoms. In order to control the influence between the two crossed nanofins, an additional term $\Delta f$ is introduced into the phase equation to modify the shape of the wavefront, whereby the phase dispersion can be easily engineered. Compared with local modulation, the number of unique nanostructures that need to be optimized can be reduced to two (operating at dual wavelengths) by the Pancharatnam-Berry (PB) phase based global modulation method.  Therefore, the proposed design strategy is expected to circumvent difficulties in the local design approaches and can find widespread applications in multiwavelength imaging and spectroscopy.
\end{abstract}


\end{frontmatter}


\section{Introduction}
Sulfur hexafluoride (SF$_6$) is commonly used as an electrical gas-insulator in high-voltage and extra-high-voltage system, including gas-insulated-switchgear and gas-insulated-transmission line, owing to its outstanding dielectric properties, which enable it to withstand high electric fields and prevent electrical arcing that could cause equipment damage \cite{Tang2017}. In cases of insulation faults, the occurrence of partial discharge and local overheating can produce SF$_6$ decomposition products, such as SO$_2$F$_2$, SOF$_2$ and other, ultimately leading to equipment damage and gas leakage \cite{Tang2017,Dong2017}. Therefore, detecting gas leaks is crucial for routine maintenance inspections in these electrical systems. The absorption peak of SF$_6$ is located at 10.6 $\upmu \rm{m}$, while one of the absorption peaks of SO$_2$F$_2$ is close to 12 $\upmu \rm{m}$ \cite{Dong2017,Zhang2022,Perkins1952,Nie2013}. Both wavelengths fall within the long-wave infrared region, which serves as a transparent window in the earth’s atmosphere \cite{Chandra2018}. Consequently, employing unmanned aerial vehicle (UAV)-based thermal cameras \cite{Bertalan2022} proves highly advantageous for efficiently conducting large-scale inspections of electric equipment during routine maintenance. Nonetheless, conventional long-wave infrared optical lenses are bulky, heavy and cumbersome \cite{Zhang2023}, which hinder the integration of thermal cameras with UAVs. 

Metasurface is a two-dimensional (2D) planar structure composed of meta-atoms or meta-molecules with sub-wavelength size, which can control the optical properties of electromagnetic waves such as amplitude, phase and polarization \cite{Yu2011,Sun2012}. Different from traditional optical devices which realize electromagnetic wavefront manipulation based on phase accumulation, the phase in metasurfaces can be accurately controlled by sub-wavelength spaced structures with thicknesses at the wavelength scale or below and thus has the advantages of being ultra-thin, ultra-light and easy integration \cite{Yu2014,Qiu2021}. Metalenses \cite{Chen2020,Chen2018,Wang2018}, optical devices based on metasurfaces, have attracted considerable interest because of their potential applications in compact imaging systems for both consumer and industry products. These innovative metalenses hold promise in addressing the challenges posed by traditional optical lenses, as mentioned above. Notably, both polarization and dispersion are key parameters in the design of metalenses \cite{Chen2020}.

Compared to polarization-dependent metalenses, polarization-insensitive metalenses are more suitable for realistic applications. Their capability to function effectively with light of any polarization allows for greater versatility and practicality. A variety of methods have been proposed to achieve polarization-insensitive functionality. A straightforward way is to select C4-symmetric nanoantennas, such as square or isotropic circle, to be meta-atoms \cite{Avayu2017,Arbabi20161,Arbabi20162,Cai2019,Wang2021,Shi2018,Dou2020}. For the anisotropic meta-atoms, polarization-insensitivity can be achieved by an area-division method \cite{Yang2017,Lin2018}, which is based on the fact that light can be decomposed into two orthogonally-polarized components. The metalens can thus be divided into two areas: one area controls the focus of the right-circularly polarized (RCP) light while the other controls the left-circularly polarized (LCP) one. In addition, some researchers have obtained polarization-insensitivity by an amplitude-division method, in which a meta-atom controls the focus of both RCP and LCP at the same time \cite{Zhang2019,Li2020}. Through this method, the ratio of the two orthogonally-polarized components can be tuned flexibly.

On the other hand, controlling the dispersion in metalens is equally significant particularly for multiwavelength applications. For highly symmetric (C4 or higher) meta-atoms, multiwavelength achromatic focusing have been realized by numerous means, including cascading multiple metalens \cite{Avayu2017}, area-division modulation based on wavelengths \cite{Arbabi20161,Arbabi20162}, multiwavelength amplitude modulation \cite{Cai2019,Wang2021}, dispersion modulation on the basis of propagation phase \cite{Shi2018,Dou2020} etc. However, all these methods have their own disadvantageous, such as bulky (cascading multiple metalens) \cite{Avayu2017}, limited efficiency (area-division and amplitude modulation methods) \cite{Arbabi20161,Arbabi20162,Cai2019,Wang2021}, and poor designing flexibility (dispersion engineering via propagation phase) \cite{Shi2018,Dou2020}. For the anisotropic meta-atoms (such as cuboids), it is also possible to realize polarization-insensitivity and dispersion controlling at multiple wavelengths concurrently. Polarization-insensitivity is achieved by rotating the cuboids by $45^{\circ}$ to induce a same phase response to the $x-$ and $y-$component of the electric field \cite{Yang2019}. In this case, dispersion engineering is realized via the propagation phase, through which the phase is tuned locally by varying the geometric parameters of the meta-atoms. However, this approach introduces numerous parameters, resulting in poor designing flexibility.

Normally, the propagation phase can be considered as a local phase modulation, while the Pancharatnam-Berry (PB) phase can be viewed as a global effect \cite{Chen2021}. Here, based on the PB phase modulation, we propose a global design strategy to circumvent the design difficulties of poor designing flexibility in dispersion engineering mentioned above. We achieve a polarization-insensitive metalens with dispersion controllability at two wavelengths of 10.6 $\upmu$m and 12 $\upmu$m (NA= 0.4). The method of amplitude-division is applied through which each meta-atom can focus both RCP and LCP simultaneously. To engineer dispersion between two design wavelengths, only the dimension of two nanofins are demanded to be optimized independently (via genetic algorithm (GA)) by utilizing our method. The two optimized nanofins are superimposed to form cross-shaped meta-atoms. The dispersion property of the resultant metalens is tuned by incorporating a wavefront shape adjustment term, $\Delta f$, into the phase equation, which can manipulate the interaction between the two nanofins. The dispersion can either be eliminated or enhanced, which leads to an achromatic metalens or a super-chromatic metalens. Compared to the local designing method, the global designing strategy proposed here is simple, timesaving and effective. Our work contributes to solving the cumbersome problem of detecting SF$_6$ gas leaks using UAV-based thermal cameras. It should be known that the design method in this paper can be extended to other bands, such as visible, terahertz and others. In addition, the polarizations (RCP and LCP) and wavelengths can be controlled independently. The multi-dimensional controllable function opens up a new door in the design of multifunctional and multiplexed metasurfaces.

\section{Theory}
For RCP/LCP illumination, if collimated incident light needs to be focused into a spot by PB phase modulation, the phase at position $(x, y)$ in the space must satisfy the following equation \cite{Zhou2020}:
\begin{equation}
\label{eq.1}
\left\{ {\begin{array}{*{20}{c}}
{{\varphi _{RCP}}\left( {x,y,\lambda } \right) =  - \frac{{2\pi }}{\lambda }\left( {\sqrt {{x^2} + {y^2} + {f^2}}  - f} \right)}\\
{{\varphi _{LCP}}\left( {x,y,\lambda } \right) = \frac{{2\pi }}{\lambda }\left( {\sqrt {{x^2} + {y^2} + {f^2}}  - f} \right)}
\end{array}} \right.,
\end{equation}
where $\lambda$ is the design wavelength and $f$ is the focal length. Since any polarized light can be decomposed into two orthogonally circularly polarized (CP) one, the phase distribution can be written as follows for the polarization-insensitive metalens based on the PB phase \cite{Zhang2019,Zhou2020}:
\begin{equation}
\label{eq.2}
\begin{array}{l}
\varphi \left( {x,y,\lambda } \right) = \arg \{ {a\exp \left[ {j{\varphi _{RCP}}\left( {x,y,\lambda } \right)} \right]} \\
\quad\quad\quad\quad~ + {b\exp \left[ {j{\varphi _{LCP}}\left( {x,y,\lambda } \right)} \right]} \} \\
\quad\quad\quad = \arg \{ {a\exp \left[ { - \frac{{2\pi j}}{\lambda }\left( {\sqrt {{x^2} + {y^2} + {f^2}}  - f} \right)} \right]}\\
\quad\quad\quad + {b\exp \left[ {\frac{{2\pi j}}{\lambda }\left( {\sqrt {{x^2} + {y^2} + {f^2}}  - f} \right)} \right]} \},
\end{array}
\end{equation}

where $a$ and $b$ are the amplitudes for RCP and LCP light, respectively. They can be tuned flexibly, depending on the required functions of the metalens. The theoretical limits of the focusing efficiency for RCP and LCP light are determined by $a/(a+b)$ and $b/(a+b)$. To realize polarization-insensitive characteristics, $a=b=1$. When illuminated by a plane wave, the transmitted electric field intensity on the metalens can be described by,
\begin{equation}
\label{eq.3}
E\left( {x,y,\lambda } \right) = B{e^{j\varphi \left( {x,y,\lambda } \right)}},
\end{equation}
where $B$ is the amplitude of the transmitted light and $j$ is the imaginary symbol of the complex number.

Under CP light illumination, the Jones matrix for an anisotropic nanofin rotated around the geometric center by an angle $\theta$ can be expressed as \cite{Lin2019}:
\begin{equation}
\label{eq.4}
J{\left( \theta  \right)_{\rm{circular}}} = \left[ {\begin{array}{*{20}{c}}
{\frac{1}{2}\left( {{t_l} + {t_s}} \right)}&{\frac{1}{2}\left( {{t_l} - {t_s}} \right){e^{j2\theta }}}\\
{\frac{1}{2}\left( {{t_l} - {t_s}} \right){e^{ - j2\theta }}}&{\frac{1}{2}\left( {{t_l} + {t_s}} \right)}
\end{array}} \right],
\end{equation}
where $t_l$ and $t_s$ are the transmission coefficients along the long and short axes of the nanofin, respectively. If two different anisotropic nanofins rotated by $\theta_1$ and $\theta_2$ respectively are superimposed together to form a cross-shaped meta-atom, its Jones matrix can be described by:
\setlength{\mathindent}{0cm}
\begin{equation}
\label{eq.5}
\begin{array}{l}
J\left( {{\theta _1},{\theta _2}} \right)_{combine} = J\left( {{\theta _1}} \right) + J\left( {{\theta _2}} \right)\\
\quad\quad = \left[ {\begin{array}{*{20}{c}}
{\frac{1}{2}\left( {{t_l} + {t_s}} \right)}&{\frac{1}{2}\left( {{t_l} - {t_s}} \right){e^{j2{\theta _1}}}}\\
{\frac{1}{2}\left( {{t_l} - {t_s}} \right){e^{ - j2{\theta _1}}}}&{\frac{1}{2}\left( {{t_l} + {t_s}} \right)}
\end{array}} \right] \\
\quad\quad  + \left[ {\begin{array}{*{20}{c}}
{\frac{1}{2}\left( {{{t'}_l} + {{t'}_s}} \right)}&{\frac{1}{2}\left( {{{t'}_l} - {{t'}_s}} \right){e^{j2{\theta _2}}}}\\
{\frac{1}{2}\left( {{{t'}_l} - {{t'}_s}} \right){e^{ - j2{\theta _2}}}}&{\frac{1}{2}\left( {{{t'}_l} + {{t'}_s}} \right)}
\end{array}} \right],
\end{array}
\end{equation}
${t'_l}$ and ${t'_s}$ represent the transmission coefficients of the second nanofin. By defining ${A_1} = {t_l} + {t_s}$, ${B_1} = {t_l} - {t_s}$, ${A_2} = {{t'}_l} + {{t'}_s}$, ${B_2} = {{t'}_l} - {{t'}_s}$, Eq. (\ref{eq.5}) can be simplified as:
\begin{equation}
\label{eq.6}
\begin{array}{l}
J{\left( {{\theta _1},{\theta _2}} \right)_{combine}} = \\
\left[ {\begin{array}{*{20}{c}}
{\frac{1}{2}\left( {{A_1} + {A_2}} \right)}&{\frac{1}{2}{B_1}{e^{j2{\theta _1}}} + \frac{1}{2}{B_2}{e^{j2{\theta _2}}}}\\
{\frac{1}{2}{B_1}{e^{ - j2{\theta _1}}} + \frac{1}{2}{B_2}{e^{ - j2{\theta _2}}}}&{\frac{1}{2}\left( {{A_1} + {A_2}} \right)}
\end{array}} \right]
\end{array}.
\end{equation}

For RCP illumination, the transmitted electric field of the cross-shaped meta-atom is:
\begin{equation}
\label{eq.7}
\begin{array}{l}
E = J{\left( {{\theta _1},{\theta _2}} \right)_{combine}}\left[ {\begin{array}{*{20}{c}}
1\\
0
\end{array}} \right]\\
 = \frac{1}{2}\left( {{A_1} + {A_2}} \right)\left[ {\begin{array}{*{20}{c}}
1\\
0
\end{array}} \right] + \frac{1}{2}{B_1}{e^{ - j2{\theta _1}}}\left[ {\begin{array}{*{20}{c}}
0\\
1
\end{array}} \right] + \frac{1}{2}{B_2}{e^{ - j2{\theta _2}}}\left[ {\begin{array}{*{20}{c}}
0\\
1
\end{array}} \right],
\end{array}
\end{equation}
$\left[ {\begin{array}{*{20}{c}}1\\0\end{array}} \right]$ and $\left[ {\begin{array}{*{20}{c}}0\\1\end{array}} \right]$ represent RCP and LCP light respectively. It can be seen that both RCP with the same phase as incident and LCP with abrupt ${2\theta_1}$ and ${2\theta_2}$ phases are included in the transmission. As only the phase of the transmitted light with an orthogonal polarization state can be modulated, for a specific wavelength $\lambda$, ${t_l}(\lambda )$ and ${t_s}(\lambda )$ of the anisotropic nanofin should have a phase difference of $\pi$ to maximize its polarization conversion efficiency (PCE), i.e., the ratio of the transmission with opposite chirality to the incident. That is, the anisotropic nanofin should meet the condition of a half-wave plate. 

Besides the polarization-insensitive characteristics, if we also want to control the dispersion property such as realizing achromatic focusing at the two operating wavelengths, we can optimize the geometric size of the two corresponding nanofins at first. They carry the phase information ${\varphi_1}$ and ${\varphi_2}$ respectively as in the following,
\begin{equation}
\label{eq.8}
\left\{ {\begin{array}{*{20}{c}}
\begin{array}{l}
{\varphi _1}\left( {x,y,{\lambda _1}} \right) =\\
 \quad\quad\arg \{ \exp \left[ { - \frac{{2\pi j}}{{{\lambda _1}}}\left( {\sqrt {{x^2} + {y^2} + {f^2}}  - f} \right)} \right]\\
 \quad\quad+ \exp \left[ {\frac{{2\pi j}}{{{\lambda _1}}}\left( {\sqrt {{x^2} + {y^2} + {f^2}}  - f} \right)} \right]\} 
\end{array}\\
\begin{array}{l}
{\varphi _2}\left( {x,y,{\lambda _2}} \right) =\\
\quad\quad \arg \{ \exp \left[ { - \frac{{2\pi j}}{{{\lambda _2}}}\left( {\sqrt {{x^2} + {y^2} + {f^2}}  - f} \right)} \right]\\
\quad\quad + \exp \left[ {\frac{{2\pi j}}{{{\lambda _2}}}\left( {\sqrt {{x^2} + {y^2} + {f^2}}  - f} \right)} \right]\} 
\end{array}
\end{array}} \right..
\end{equation}

After the two optimized nanofins are superimposed to form cross-shaped meta-atoms, the transmitted electric field can be expressed as,
\begin{equation}
\label{eq.9}
E\left( {x,y,\lambda } \right) = {B_1}\left( \lambda  \right){e^{j{\varphi _1}\left( {x,y,{\lambda _1}} \right)}} + {B_2}\left( \lambda  \right){e^{j{\varphi _2}\left( {x,y,{\lambda _2}} \right)}},
\end{equation}
where, ${B_1(\lambda)}$ and ${B_2(\lambda)}$ are the transmitted light amplitudes of the two nanofins at a wavelength of $\lambda$ respectively.

Ideally, if ${B_1}({\lambda _1}):{B_2}({\lambda _1}) = 1:0$ for ${\lambda _1}$ incident while ${B_1}({\lambda _2}):{B_2}({\lambda _2}) = 0:1$ for ${\lambda _2}$, the shape of the wavefronts and the focal length should be identical at both wavelengths and the system is achromatic. In practice, however, it is impossible to find ideal nanofins to meet the requirements. As shown in Eqs. (\ref{eq.7}) and (\ref{eq.9}), one of the nanofins controls the phase of its design wavelength, the phase modulation from the other nanofin becomes an interference term. In order to minimize it, the condition ${{B_1}({\lambda _1}) \gg {B_2}({\lambda _1})({B_2}({\lambda _2}) \gg {B_1}({\lambda _2}))}$  should be satisfied for ${\lambda _1}({\lambda _2})$ incident.

\section{Achromatic and super-chromatic focusing}
\begin{figure}[htbp]
\centering
\includegraphics[width=\linewidth]{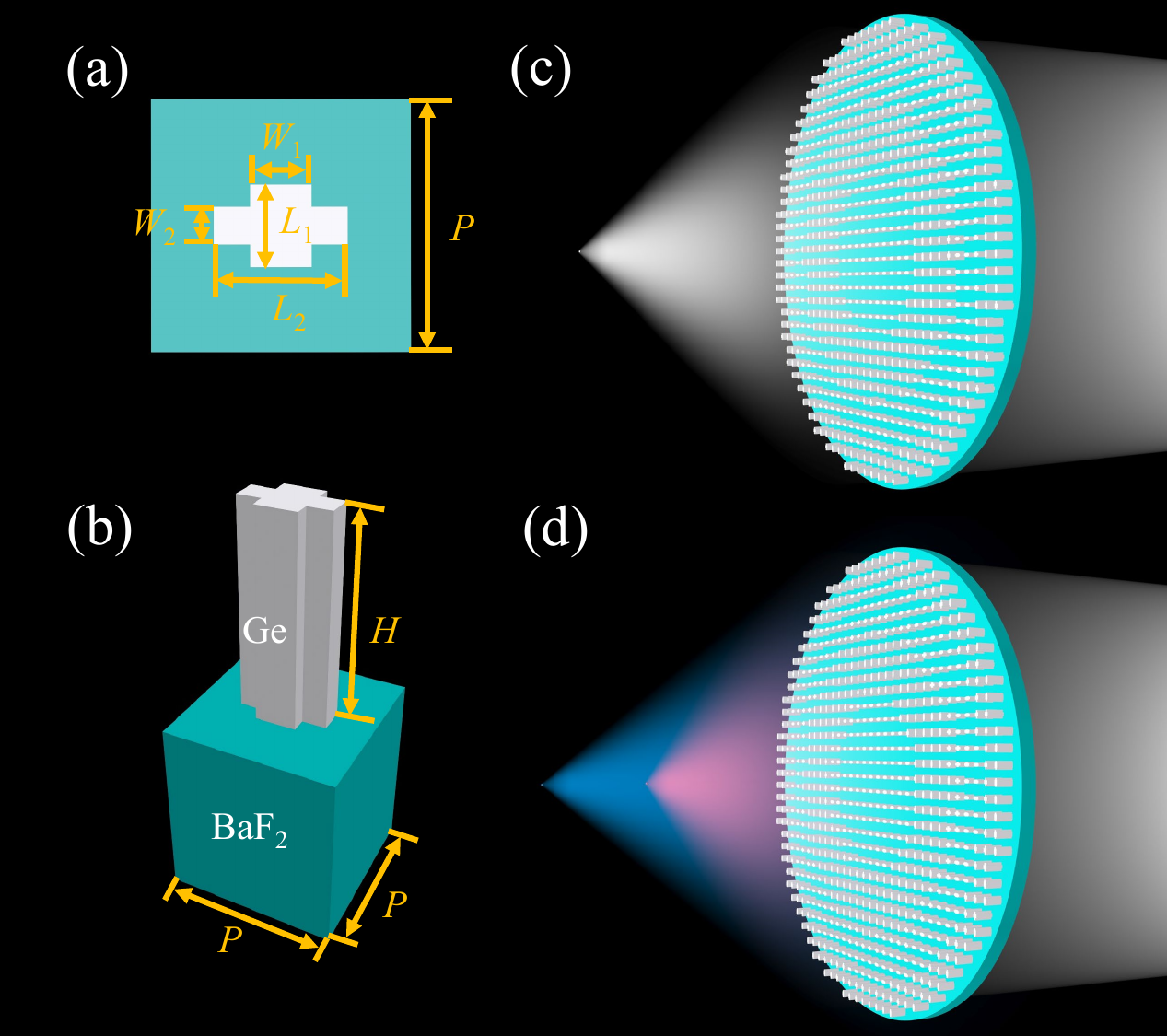}
\caption{(a)/(b) The top/side view of the cross-shaped meta-atom. The meta-atom height ${H=6.8~\upmu \rm{m}}$, period  ${P=6~\upmu \rm{m}}$; (c)/(d) The schematic diagram of the achromatic/super-chromatic metalens.}
\label{fig.1}
\end{figure}

In the numerical simulations, the building blocks of the metalens are chosen to be Ge nanofins \cite{Lcenogle1976} on a $\rm {BaF_2}$ substrate \cite{Li1980}. We set ${\lambda_1=10.6~ \upmu \rm{m}}$ and ${\lambda_2=12~ \upmu \rm{m}}$ to demonstrate the design strategy here. The design strategy can be easily extended to other wavebands, including those in visible. Corresponding to $\lambda_1/\lambda_2$, the geometric parameters of the nanofins are optimized (by GA) to be: length $L_1/L_2=1.96/3.10~\upmu \rm{m}$, width $W_1/W_2=1.42/0.90~\upmu \rm{m}$. They are crossed with each other as shown in Fig.\ref{fig.1}(a) and (b). The period and height of the nanofin are set at $P=6~\upmu \rm{m}$ and $H=6.8~\upmu \rm{m}$, respectively, which can be achieved by a state-of-the–art standard nanofabrication technology \cite{Liu2019}. At ${\lambda_1}$, the PCE of nanofin 1 and 2 are 99.1\% and 4.4\% which results in ${B_1}({\lambda _1}):{B_2}({\lambda _1}) = 22.5:1$. At ${\lambda_2}$, ${B_1}({\lambda _2}):{B_2}({\lambda _2}) = 1:10.3$ as the PCE are 9.1\% and 93.6\% respectively. Therefore, we can consider that the PCE of the optimized meta-atoms satisfies the requirements for  ${{B_1}({\lambda _1}) \gg {B_2}({\lambda _1})({B_2}({\lambda _2}) \gg {B_1}({\lambda _2}))}$ at ${\lambda _1}({\lambda _2})$. The finite-difference time-domain (FDTD) method (FDTD Solutions, Lumerical, Canada) is used for numerical calculations here. The mesh size 150 nm was found to be small enough to gain converging results. In order to satisfy the Nyquist sampling criterion \cite{Chen2020,Chen2021} to prevent additional diffraction orders, the numerical aperture (NA) of the metalens should be smaller than ${\lambda /2P}$. To increase fault tolerance, the NA is set at 0.4. Considering the limitations of computer memory during simulation, the radius of the metalens $R$ is set to be $96~\upmu \rm{m}$ and the focal length $f$ is set to be $220~\upmu \rm{m}$. The focusing efficiency is defined as the ratio of the power at the focal spot to the power of incident.

\begin{figure*}[htbp]
\centering
\includegraphics[width=15cm]{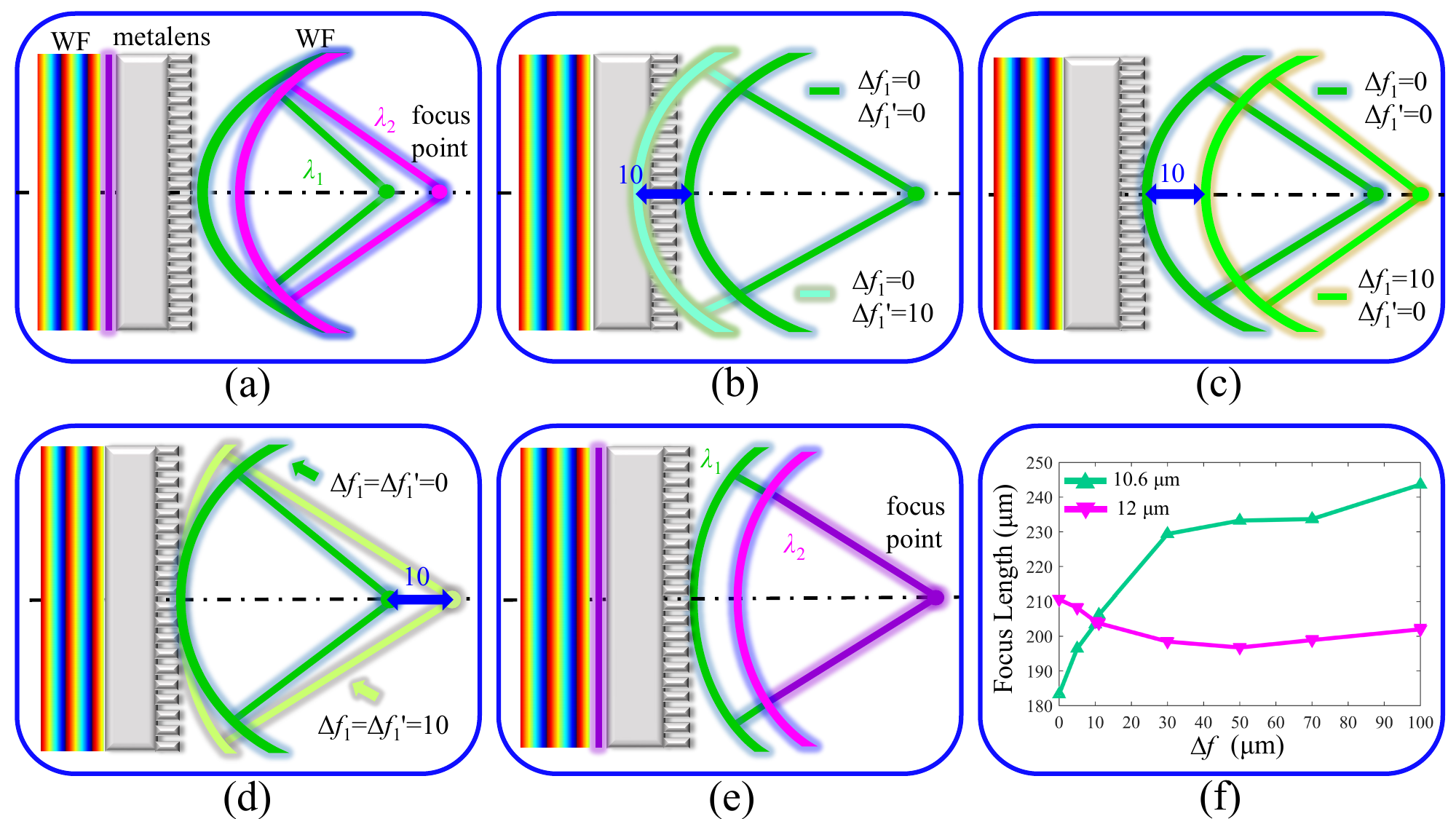}
\caption{ Schematic of the wavefronts before and after the metalens. (a) Focusing at ${\lambda _1}$ and ${\lambda _2}$ with chromatic characteristics; (b)-(d) The variation of the wavefront shape with a change of $\Delta {f_1}$ and $\Delta {f'_1}$ comparing with $\Delta {f_1}=\Delta {f'_1}=0$ at ${\lambda _1}$. WF: wavefront. (e) Focusing at ${\lambda _1}$ and ${\lambda _2}$ achromatically after optimization. (f) The influence of $\Delta {f}$ on the focal length with RCP illumination.}
\label{fig.2}
\end{figure*}

\begin{figure*}[htbp]
\centering
\includegraphics[width=18cm]{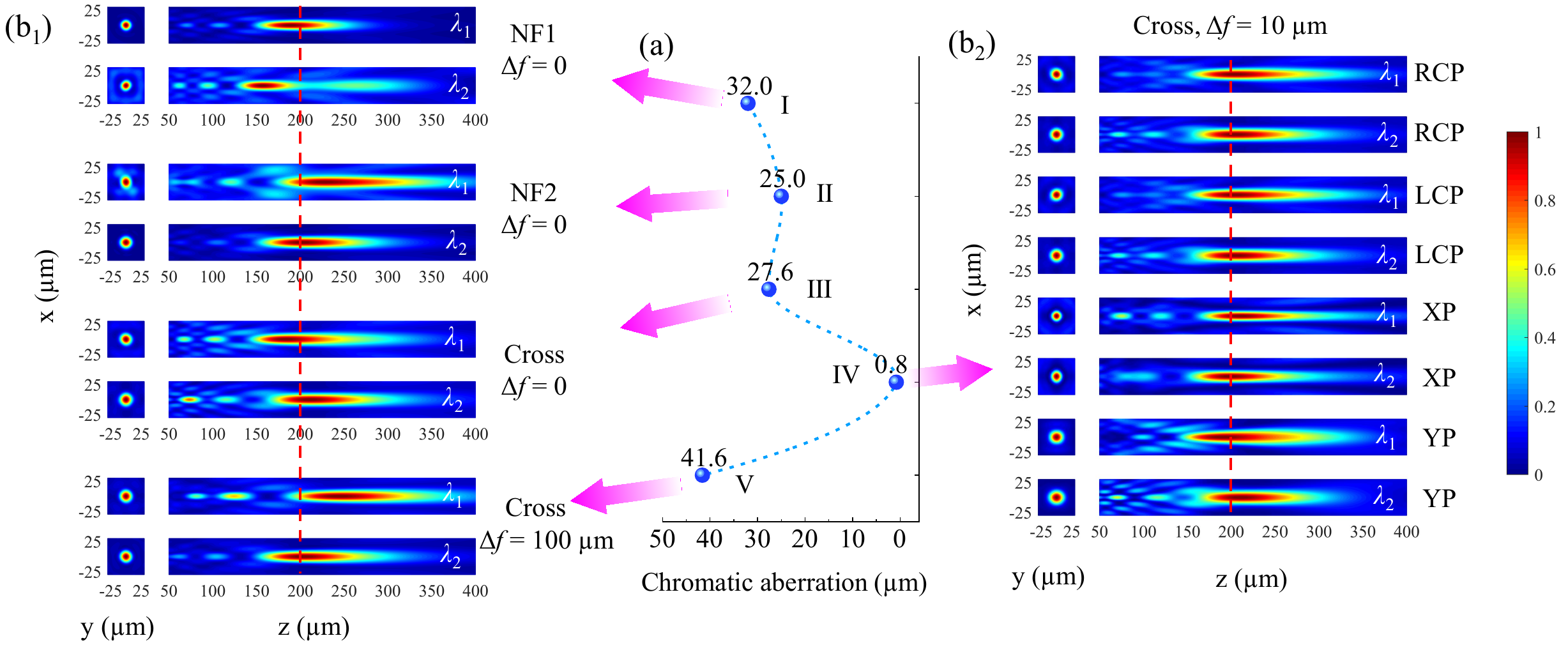}
\caption{(a) The difference of the focal length between $\lambda_1$ and $\lambda_2$ under different conditions. Nanofin: NF. Case I/II: the meta-atom is NF1/NF2, $\Delta {f}=0$, RCP input; Case III/V: cross-shaped meta-atom, $\Delta {f}=0/100~\upmu \rm{m}$, RCP input; Case IV: cross-shaped meta-atom, $\Delta {f}=10~\upmu \rm{m}$, different polarizations. $(\rm{b_1})/(\rm{b_2})$ Corresponding intensity profiles in the $x-y$ and $x-z$ plane. $\lambda_1=10.6~\upmu \rm{m}$, $\lambda_2=12~\upmu \rm{m}$. The red dashed lines indicate the position at $z=200~\upmu \rm{m}$.}
\label{fig.3}
\end{figure*}

If the two geometrical-parameter-optimized nanofins are directly crossed together, they will interfere each other. This will lead to the wavefronts at the two wavelengths being different and cause chromatic aberration, as schematically shown in Fig. \ref{fig.2}(a). The corresponding FDTD simulation results in Fig. \ref{fig.3}(a) show that the actual focal length at ${\lambda _1}/{\lambda _2}$ are $183.2/210.8~\upmu \rm{m}$ respectively, bringing about a chromatic aberration $27.6~\upmu \rm{m}$ (Case III). This value is similar to those obtained from the two metalens composed of nanofins optimized for ${\lambda _1}$ ($32.0~\upmu \rm{m}$, Case I) and ${\lambda _2}$ ($25.0~\upmu \rm{m}$, Case II) severally.

As the two nanofins used to form cross-shaped meta-atoms are designed to control their own phases independently, the interaction between them should be able to be compensated by adjusting one of the wavefronts. To make it simple, the metalens is considered to be polarization-sensitive and the compensation part is only added onto the wavefront of ${\lambda _1}$, thus the phase $\varphi_1$ can be expressed as in the following,
\begin{equation}
\label{eq.10}
\varphi_1  =  - \frac{{2\pi }}{{{\lambda _1}}}\left( {\sqrt {{x^2} + {y^2} + {{(f + \Delta {f_1})}^2}}  - (f + \Delta {f'_1})} \right).
\end{equation}
Obviously, both $\Delta {f_1}$ and $\Delta {f'_1}$  can influence the phase. The compensation part is incorporated into the global phase of the metalens which are realized by rotating the nanofins. Optimized Nanofin 1 (corresponding to $\lambda _1$) is applied in the metalens. Without loss of generality, we take $\Delta {f_1}/\Delta {f'_1}$ as 0 and 10 $\upmu \rm{m}$ in the FDTD simulations to investigate their influence.

\begin{table}[htbp]
\setlength\tabcolsep{3pt}
\centering
\caption{FDTD simulation results for RCP incident at a wavelength of $10.6~\upmu \rm{m}$. Geometrical parameters of the Nanofin 1: ${L_1=1.96~\upmu \rm{m}}$, ${W_1=1.42~\upmu \rm{m}}$. FL: Focus Length. FWHM: full width at half maximum. }
\begin{tabular}{ccccc}
\hline
    {$\Delta f~(\upmu \rm{m})$} & Figure  & FL & {FWHM} & {FWHM}\\
      {$(\Delta f_1, \Delta f'_1)$} &  Reference   &   $(\upmu \rm{m})$   &  $-x$ $(\lambda)$  &  $-y$ $(\lambda)$ \\
\hline
{$(0,~0)$} & Fig. \ref{fig.2}(b)-(d) &{${199.4}$} & 1.17 & 1.22 \\
{$(0,~10)$} & Fig. \ref{fig.2}(b) &{${199.8}$} & 1.18 & 1.24 \\ 
{$(10,~0)$} & Fig. \ref{fig.2}(c)  &{${207.3}$} & 1.25 & 1.27 \\
{$(10,~10)$} & Fig. \ref{fig.2}(d)  &{${207.7}$} & 1.22 & 1.27\\ 
\hline
\end{tabular}
  \label{table.1}
\end{table}

From Eq. (\ref{eq.10}), we can derive that the incorporation of $\Delta {f'_1}$ can only shift the whole wavefront forward or backward but cannot affect its shape, thus the focal length can be maintained (refer to Fig. \ref{fig.2} (b)). This point can be confirmed by comparing the FDTD simulation results shown in the first two rows in Table \ref{table.1}. For $\Delta {f_1}$, based on its influence on the phase in Eq. (\ref{eq.10}), it can change the focal length by influencing the shape of the wavefront (refer to Fig. \ref{fig.2} (c) and (d)), which are accordant with the results shown in the $1^{\rm{st}}$, $3^{\rm{rd}}$ and $4^{\rm{th}}$ rows of Table \ref{table.1}. When $\Delta {f_1}=\Delta {f'_1}$, the phase $\varphi_1$ at $x=y=0$ is the same, thus the apex position of the wavefront overlaps (refer to Fig. \ref{fig.2} (d)).

From Table \ref{table.1}, it can be seen that $\Delta {f_1}$ is the key parameter to determine the focal length. When $\Delta {f_1}$ is the same, the focal length and the wavefront is maintained whether $\Delta {f'_1}$ is changed or not. Therefore, we can set $\Delta {f}=\Delta {f_1}=\Delta {f'_1}$ to unify the calculation in Eq. (\ref{eq.10}).

As discussed above, by adding a proper $\Delta {f}$, we can tune one wavefront while affect the other one. Then the two wavefronts can be adjusted to focus at the same point, as shown in Fig. \ref{fig.2}(e). However, we have to note that when two nanofins are crossed onto each other, the interaction between them cannot be defined accurately. Therefore, we have to scan the value of $\Delta {f}$ to optimize its effect on the wavefront. We can choose to tune the wavefront of $\lambda_1$ or $\lambda_2$ to realize achromatic or super-chromatic functions. Without loss of generality, we still select $\lambda_1$ to investigate and add $\Delta {f}$ into its phase. According to Eq. (\ref{eq.8}), for polarization-insensitive metalens, the phase at $\lambda_1$ and $\lambda_2$ can be expressed as,

\begin{equation}
\label{eq.11}
\left\{ {\begin{array}{*{20}{c}}
\begin{array}{l}
{\varphi _1}\left( {x,y,{\lambda _1}} \right) = \\
 \arg \{ \exp \left[ { - \frac{{2\pi j}}{{{\lambda _1}}}\left( {\sqrt {{x^2} + {y^2} + {{\left( {f + \Delta f} \right)}^2}}  - \left( {f + \Delta f} \right)} \right)} \right]\\
~ + \exp \left[ {\frac{{2\pi j}}{{{\lambda _1}}}\left( {\sqrt {{x^2} + {y^2} + {{\left( {f + \Delta f} \right)}^2}}  - \left( {f + \Delta f} \right)} \right)} \right]\} 
\end{array}\\
\begin{array}{l}
{\varphi _2}\left( {x,y,{\lambda _2}} \right) =\quad\quad\quad\quad\quad\quad\quad\quad\quad\quad\quad\quad\quad\quad\quad\quad\quad\quad \\
 \arg \{ \exp \left[ { - \frac{{2\pi j}}{{{\lambda _2}}}\left( {\sqrt {{x^2} + {y^2} + {f^2}}  - f} \right)} \right]\\
~ + \exp \left[ {\frac{{2\pi j}}{{{\lambda _2}}}\left( {\sqrt {{x^2} + {y^2} + {f^2}}  - f} \right)} \right]\} 
\end{array}
\end{array}} \right..
\end{equation}

Fig. \ref{fig.2} (f) demonstrates that, the focal length $f_1$ at $\lambda_1$ increases significantly/ gradually with $\Delta {f}$ when $\Delta f \le 30~\upmu \rm{m}/\Delta f > 30~\upmu \rm{m}$. Although $\Delta {f}$ is added into the phase at$\lambda_1$, $f_2$ at $\lambda_2$ also varies with $\Delta {f}$. This is due to the interference between the two crossed nanofins . The numerical results show that when $\Delta {f}=10~\upmu \rm{m}$, the focal length at$\lambda_1$ and $\lambda_2$ are $203.5~\upmu \rm{m}$ and $204.3~\upmu \rm{m}$, respectively. The chromatic aberration is thus reduced to only $0.8~\upmu \rm{m}$, indicating that the metalens is basically achromatic at the wavelength of $\lambda_1$ and $\lambda_2$. Furthermore, the focal spots are symmetric (Fig. \ref{fig.3}$(\rm{b_2})$) with FWHMs similar to the theoretical diffraction limited value $\lambda / 2\rm{NA}=1.25\lambda$  for all the polarizations at both wavelengths (Table \ref{table.2}).

\begin{table}[htbp]
\setlength\tabcolsep{2pt}
\centering
\caption{The detailed results for the designed metalens with $R=96~\upmu \rm{m}$ and $f=220~\upmu \rm{m}$, giving NA=0.4. $\Delta {f}=10~\upmu \rm{m}$, $\lambda_1/\lambda_2=10.6/12~\upmu \rm{m}$. FL: Focus length. CA: Chromatic aberration. FE: Focusing efficiency}
\resizebox{0.49\textwidth}{!}{
\begin{tabular}{ccccccccc}
\hline
   Polarization   & \multicolumn{2}{c}{RCP} & \multicolumn{2}{c}{LCP} & \multicolumn{2}{c}{XP} & \multicolumn{2}{c}{YP}  \\
      $\lambda$ ($\upmu \rm{m}$)  & {10.6}  & {12}  &  {10.6}  & {12}  & {10.6}  & {12}  & {10.6}  & {12}  \\
\hline
  FL ($\upmu \rm{m}$)  & {203.5}  & {204.3}  &  {203.5}  & {204.3}  & {209.5}  & {198.5}  & {190.3}  & {209.9}  \\
  CA ($\upmu \rm{m}$) & \multicolumn{2}{c}{0.8} & \multicolumn{2}{c}{0.8 } & \multicolumn{2}{c}{11.0 } & \multicolumn{2}{c}{19.6}  \\
     FWHM-$x$  & \multirow{2}{*}{1.37}  & \multirow{2}{*}{1.30}  &  \multirow{2}{*}{1.37}  & \multirow{2}{*}{1.30}  & \multirow{2}{*}{1.17} & \multirow{2}{*}{1.17}  & \multirow{2}{*}{1.50}  & \multirow{2}{*}{1.44}  \\
  $(\lambda)$ & & & & & & & &  \\
     FWHM-$y$  & \multirow{2}{*}{1.37}  & \multirow{2}{*}{1.28}  &  \multirow{2}{*}{1.37}  & \multirow{2}{*}{1.28}  & \multirow{2}{*}{1.12} & \multirow{2}{*}{1.06}  & \multirow{2}{*}{1.63}  & \multirow{2}{*}{1.61}  \\
  $(\lambda)$ & & & & & & & &  \\
  FE (\%) & 35.3  & 35.0 &  35.3 & 35.0  & 29.9  & 32.8  & 45.8  & 43.8  \\
\hline
\end{tabular}
}
  \label{table.2}
\end{table}

To demonstrate the polarization-insensitivity of the designed metalens, we calculate the intensity profile at the focus in the $x-z$ and $x-y$ plane with different polarizations as shown in Fig. \ref{fig.3}$(\rm{b_2})$. The detailed results are given in Table \ref{table.2}. It can be seen that the focal length and FWHM of the focus spot of the metalens are basically the same under RCP and LCP illuminations. However, the difference between $X-$ and $Y-$polarization (XP and YP) cannot be neglected.

\begin{figure}[htbp]
\centering
\includegraphics[width=\linewidth]{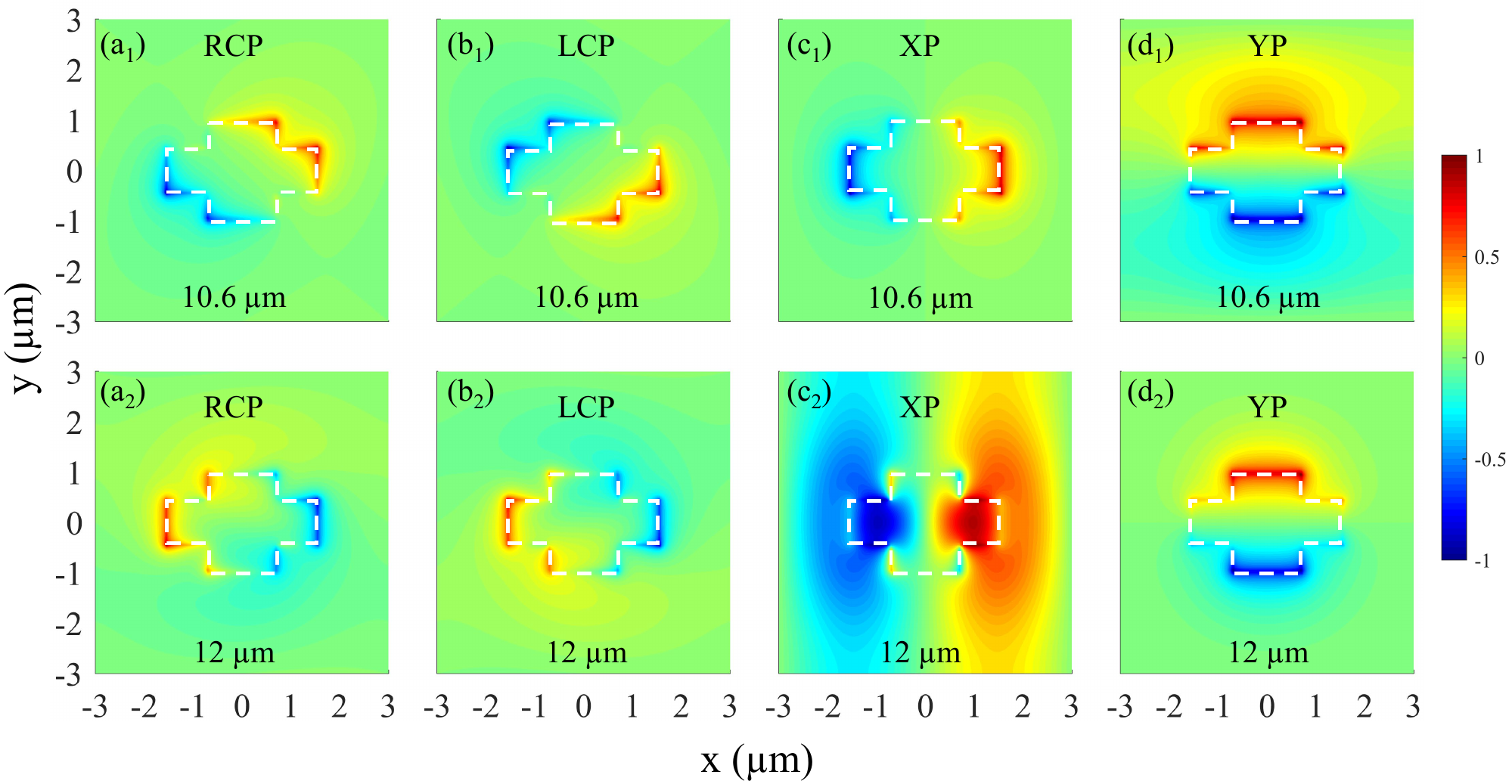}
\caption{The distribution of the $z$ component of the normalized electric field on the surface of the cross-shaped meta-atom at wavelengths of 10.6 (top panel) and 12 $\upmu \rm{m}$ (bottom panel) under different polarizations. The white dashed lines mark the boundaries of the meta-atom.}
\label{fig.4}
\end{figure}

To gain insight into the origin of the difference, we then calculate the $z$-component of the electric field ($E_z$) along the surface of a cross-shaped meta-atom under different wavelengths and different polarizations. As shown in Fig. \ref{fig.4}, the incident light excites the dipole radiation in the meta-atom, in which the direction of the dipole moment is related with the polarization of the excitation. For RCP and LCP incident light, the distribution of $E_z$ is identical except the opposite direction (Fig. \ref{fig.4}(a) and (b)). This can be attributed to the identical dipole radiation response of the meta-atom. When the incident is linearly polarized along $x-$ and $y-$direction, however, the dipole radiation response and thus the $E_z$ distribution is different both at $\lambda_1$ and $\lambda_2$ (Fig. \ref{fig.4}(c) and (d)). As a result, we can observe that the intensity profiles either on $x-z$ or $x-y$ plane are slightly different (Fig. \ref{fig.3}($\rm{b_2}$)). Not surprise, the focal lengths for $x-$ and $y-$polarization are not the same and they are also different from that for LCP or RCP incident. Although the chromatic aberration between 10.6 and 12 $\upmu \rm{m}$ is not negligible both for $x-$ and $y-$polarization, they are however much smaller than the focal depth, which equals to $\lambda/\rm{NA}^2 = 10.6~\upmu \rm{m}/0.4^2 = 66.3~\upmu \rm{m}$. They are also smaller than those values obtained in Case I, II, and III (refer to Fig. \ref{fig.3}(a)).

As the wavefront shape can be adjusted by $\Delta {f}$ according to Eq. (\ref{eq.11}), besides being able to reduce or even eliminate the achromatic aberration, we can also realize super-chromatic function in the designed polarization-insensitive metalens. From Fig. \ref{fig.2}(f), it can be observed that the difference of the focal length increases slowly with $\Delta {f}$ (when $\Delta f > 30~\upmu \rm{m}$) and becomes stable at last, indicating that the dispersion is limited somehow. When $\Delta {f}=100~\upmu \rm{m}$, the focal length at the wavelengths of $10.6~\upmu \rm{m}$ and $12~~\upmu \rm{m}$ are $243.6~~\upmu \rm{m}$ and $202.0~\upmu \rm{m}$, respectively. In this case, the chromatic aberration reaches $41.6~\upmu \rm{m}$ (corresponding to the case V in Fig. \ref{fig.3} (a) and $(\rm{b_1})$). A schematic of the metalens based on the super-chromatic property is shown in Fig. \ref{fig.1} (d). 

At last, it is worth to note the limitation of the focusing efficiency of the designed metalens. According to Eq. (\ref{eq.2}) and the mechanism of the PB phase, the polarization-insensitive function can only be realized by rotating the anisotropic nanofin with $0^{\circ}$ and $90^{\circ}$. They compensate a phase of 0/0 or $\pi/(-\pi)$ for RCP/LCP incident. As all the nanofins only have two rotation angles, the entire metalens only consists of two phase gradients, i.e. 0 and $\pi$. Similar to the binary phase Fresnel zone plate \cite{Li2020}, the metalens has a real focal point and simultaneously a diverging virtual focal point, as well as several secondary focal points with lower intensity. Therefore, the focusing efficiency is limited to be 50\% theoretically. From our simulation results, the focusing efficiency reaches to 35\% and this value is comparable to other reported metalens consisting of isotropic meta-atoms to achieve polarization-insensitive characteristics \cite{Avayu2017,Arbabi20161,Arbabi20162,Cai2019,Wang2021,Shi2018,Dou2020}.

\section{Conclusion}
In summary, in order to circumvent the disadvantage of the local designing method, in which numerous meta-atoms are required to be optimized for dual–wavelength manipulation, we propose a novel strategy to globally design a metalens with characteristics of polarization-insensitivity and dispersion-controllability at two operating wavelengths by only optimizing the dimensions of two nanofins. The meta-atom of the metalens consists of the two optimized nanofins crossed with each other. The interaction between the two nanofins is controlled by introducing $\Delta {f}$ into the phase equation. The design method is expected to find widespread applications in multiwavelength imaging and spectroscopy. Furthermore, it can be extended to other optical devices that need to modulate multiwavelength and polarization properties at the same time, such as holograms, vortex beam generators, beam shapers, and so forth.

\section*{Declaration of Competing Interest}
The authors declare that they have no known competing financial interests or personal relationships that could have appeared to influence the work reported in this paper.

\section*{Acknowledgement}
This work was funded by National Natural Science Foundation of China (NSFC) (Grant No. 12274148) and Natural Science Foundation of Guangdong province (Grant No. 2021A1515010286).

\bibliographystyle{elsarticle-num}
\bibliography{Chen-Wu-IPT}
\end{document}